\documentstyle[11pt,ihepconf,twoside,epsfig]{article}

\def\<{\langle}
\def\>{\rangle}
\newcommand{\be}{\begin{equation}}
\newcommand{\ee}{\end{equation}}
\newcommand{\bn}{\begin{eqnarray}}
\newcommand{\en}{\end{eqnarray}}
\newcommand{\bnn}{\begin{eqnarray*}}
\newcommand{\enn}{\end{eqnarray*}}
\newcommand{\ba}{\begin{array}}
\newcommand{\ea}{\end{array}}

\newcommand{\ben}{\begin{enumerate}}
\newcommand{\een}{\end{enumerate}}

\newcommand{\eti}{{\eta}_i}

\newcommand{\etiz}{{\eta}_i^0}
\newcommand{\etit}{{\eta}_i^3}

\newcommand{\la}{\lambda }

\newcommand{\La}{\Lambda }
\newcommand{\Lac}{{\Lambda}_c }

\newcommand{\ksi}{{\xi}_i}

\newcommand{\nui}{{\nu}_i}

\newcommand{\psx}{\psi\,(x)}
\newcommand{\tpsx}{\tilde \psi\,(x)}

\newcommand{\tpsmx}{\tilde \psi\,(- x)}
\newcommand{\tpsy}{\tilde \psi\,(y)}
\newcommand{\tpscy}{\tilde \psi^+\,(y)}
\newcommand{\tpsyj}{\tilde \psi\,(y_j)}

\newcommand{\tpsox}{\tilde \psi_1\,(x)}
\newcommand{\tpsdx}{\tilde \psi_2\,(x)}

\newcommand{\Psz}{\Psi_0}
\newcommand{\Pszl}{\langle \Psi_0}
\newcommand{\Pszr}{\Psi_0 \rangle}

\newcommand{\vf}{\varphi }
\newcommand{\vfx}{\varphi\,(x) }
\newcommand{\vfhx}{\varphi\,(\hat{x}) }
\newcommand{\vfxo}{\varphi\,(x_1) }
\newcommand{\vfxd}{\varphi\,(x_2) }

\newcommand{\vfxn}{\varphi\,(x_n) }
\newcommand{\vfxip}{\vfxo\,\vfxd\,...\,\vfxn\,}
\newcommand{\vfxipts}{\vfxo\,\ts\,\vfxd\,\ts...\,\vfxn\,}

\newcommand{\vfhy}{\varphi\,(\hat y) }

\newcommand{\vff}{\varphi_f }
\newcommand{\tvf}{\tilde \varphi }
\newcommand{\tvfcx}{\tilde \varphi^+\,(x) }
\newcommand{\tvfcy}{\tilde \varphi^+\,(y) }

\newcommand{\tvfx}{\tilde \varphi\,(x) }
\newcommand{\tvfxo}{\tilde \varphi\,(x_1) }
\newcommand{\tvfxi}{\tilde \varphi\,(x_i) }
\newcommand{\tvfxn}{\tilde \varphi\,(x_n) }
\newcommand{\tvfmx}{\tilde \varphi\,(- x) }
\newcommand{\tvfxipo}{\tilde \varphi\,(x_{i +1}) }
\newcommand{\tvfxipopla}{\tilde \varphi\,(x_{i +1} + \la\,a) }
\newcommand{\tvfxnpla}{\tilde \varphi\,(x_n + \la\,a) }
\newcommand{\tvfy}{\tilde \varphi\,(y) }

\newcommand{\tvftxo}{\tilde \varphi\,(\tilde x_1) }
\newcommand{\tvftxi}{\tilde \varphi\,(\tilde x_i) }
\newcommand{\tvftxn}{\tilde \varphi\,(\tilde x_n) }
\newcommand{\tvftxipo}{\tilde \varphi\,(\tilde x_{i +1}) }

\newcommand{\tvfmxo}{\tilde \varphi\,(- x_1) }

\newcommand{\tvfmxi}{\tilde \varphi\,(- x_i) }
\newcommand{\tvfmxipo}{\tilde \varphi\,(- x_{i +1}) }
\newcommand{\tvfmxn}{\tilde \varphi\,(- x_n) }

\newcommand{\tvfxp}{\tilde \varphi \,(x_1) \cdots \tilde
\varphi\,(x_n) }

\newcommand{\tvfmxpo}{\tilde \varphi \,(- x_n) \cdots \tilde
\varphi\,(- x_1) }

\newcommand{\Th}{\Theta}

\def\thzi{\theta_{0i}}
\def\thod{\theta_{12}}
\def\thdo{\theta_{21}}

\newcommand{\xz}{x_0}
\newcommand{\xo}{x_1}
\newcommand{\xd}{x_2}
\newcommand{\xt}{x_3}
\newcommand{\xit}{x_i}
\newcommand{\xn}{x_n}

\newcommand{\txo}{\tilde x_1}
\newcommand{\txd}{\tilde x_2}
\newcommand{\txit}{\tilde x_i}
\newcommand{\txn}{\tilde x_n}
\newcommand{\txj}{\tilde x_j}
\newcommand{\txitz}{\tilde x_i^0}
\newcommand{\txjz}{\tilde x_j^0}
\newcommand{\txitt}{\tilde x_i^3}
\newcommand{\txjt}{\tilde x_j^3}
\newcommand{\txipo}{\tilde x_{i + 1}}

\newcommand{\yi}{y_i}

\newcommand{\zi}{z_i}

\newcommand{\wf}{W\,(x_1, x_2,..., x_n)}

\newcommand{\wfnx}{W\,(\nu_1,   \ldots, \nu_{n - 1}, X)}
\newcommand{\wftx}{W\,(\tilde x_1,...\tilde x_n)}

\newcommand{\wftxi}{W\,(\tilde x_1,   \ldots, \tilde x_i, \tilde x_{i
+1},
\ldots, \tilde x_n)}
\newcommand{\wftxio}{W\,(\tilde x_1,   \ldots, \tilde x_{i + 1},
\tilde x_i,
\ldots, \tilde x_n)}
\newcommand{\wfz}{W\,(z_1,..., z_n)}
\newcommand{\wfztx}{W\,(z_1,...z_{i - 1}, \tilde x_i, \tilde x_{i +
1},
z_{i + 2},..., z_n)}
\newcommand{\wfztxo}{W\,(z_1,...z_{i - 1}, \tilde x_{i + 1}, \tilde
x_i,
z_{i + 2},..., z_n)}
\newcommand{\wfzp}{W\,(z'_1,...z'_i, z'_{i + 1},..., z'_n)}
\newcommand{\wfzpo}{W\,(z'_1,...z'_{i + 1}, z'_i,..., z'_n)}

\newcommand{\pitx}{\Pi\,(\tilde x_1,...\tilde x_n)}
\newcommand{\wfksx}{W\,(\xi_1,   \ldots, \xi_{n - 1}, X)}

\newcommand{\ts}{\tilde \star }

\newcommand{\cpn}{P_n}
\newcommand{\cpnz}{P_n^0}
\newcommand{\cpnt}{P_n^3}
\newcommand{\vcpn}{\vec{P_n}}

\newcommand{\ctn}{T_n}
\newcommand{\ctnmi}{T_n^-}

\newcommand{\fsx}{f\,(x)}

\newcommand{\complessi}{\hbox{\rm I\hskip-5.9pt\bf C}}

\newcommand{\identity}{\hbox{\rm I\hskip-5.9pt\bf I}}

\newcommand{\cc}{commutative case }
\newcommand{\nc}{noncommutative }

\newcommand{\con}{condition }

\begin{document}

\setcounter{page}{3}
\begin{center}

{\Large \bf  RIGOROUS RESULTS in SPACE-SPACE NONCOMMUTATIVE}

\vspace*{0.2cm}
{\Large\bf QUANTUM
FIELD THEORY }\\[2.9mm]

{\large\bf M.N.~ Mnatsakanova}\footnote{e-mail:
mnatsak@theory.sinp.msu.ru} 

{\it Skobeltsyn Institute of
Nuclear 
Physics, Moscow State University, Moscow, Russia}

\vspace{0.1cm}

{\large\bf Yu.S. Vernov}\footnote{e-mail:
vernov@ms2.inr.ac.ru} 

{\it Institute
for Nuclear Research, RAS, Moscow, Russia}

\end{center}

\vspace*{0,5cm}
\begin{abstract}
The axiomatic approach based on Wightman functions is developed in
noncommutative quantum field theory. We have proved that the main
results of the axiomatic approach remain valid if the
noncommutativity affects only the spatial variables.
\end{abstract}
\vspace*{5mm}

\section{Introduction}

The axiomatic approach developed in the works by Wightman, Jost,
Bogoliubov, Haag, and others (see [1]--[4] and the references
therein) allowed stating quantum field theory as a consistent
mathematically rigorous science. This approach was first
generalized to gauge field theory in the works by Morchio and
Strocchi [5], [6].

In the framework of this approach, some fundamental results were
obtained based on general principles of the theory, for example,
the CPT theorem and the spin–statistics theorem were proved.
Analytic properties of scattering amplitudes with respect to
energy and transferred momentum were established, based on which
rigorous bounds on the high-energy behavior of the amplitudes were
derived (see [7]).

Widespread consideration is currently given to various
generalizations of the standard quantum field theory, in
particular, to noncommutative quantum field theory, whose
foundation includes the coordinate commutation relations
\be\label{cr} [\hat{x}_\mu,\hat{x}_\nu]=i\,\theta_{\mu\nu}, \ee
where $\theta_{\mu\nu}$ is a constant matrix. Interest in this kind
of theory has a long history [8]. A new developmental stage is
related, on one hand, to the construction of noncommutative
geometry [9] and, on the other hand, to new arguments in favor of
this generalization of the theory to ultrashort distances, i.e.,
to ultrahigh energies [10]. Furthermore, there is great interest
in these theories because they are low-energy limits of string
theories in some cases [11] (see [12], [13] for a survey of the
main results in this research direction). The noncommutative
theories naturally decompose into two classes, namely, those in
which time commutes with the spatial variables, i.e. $\theta_{0i}
= 0$, and those (this is the general case) in which $\theta_{0i}
\neq 0$. If  $\theta_{0i} = 0$, then the theory involves no
difficulties related to unitarity and causality, which are
characteristic of the general case [14]--[16], although there are
also some versions of a consistent theory in the general case
[17]. We note that the version with $\theta_{0i} = 0$ is precisely
the low-energy limit of string theory [11].

The major part of calculations in noncommutative field theory is
connected with perturbation theory. But a more general approach
was developed in some works (see [18]--[29]). In particular, the
investigations in [24], [25] have their origin in the ideas of
algebraic field theory, and the consideration in [26]--[29] is
based on introducing Wightman functions. The present report is
devoted to the development of the approach in [26]--[29].

As in [26]--[29], we consider the version of the theory in which
$\theta_{0i} = 0$. In this case, without loss of generality [30],
we can choose a representation that determines the algebra defined
by relation (1) and for which only the condition $\thod = - \thdo
\neq 0$ holds. Consequently, we have two noncommutative
coordinates,  $\xo$ è $\xd$, and two commutative ones,  $\xz$ è
$\xt$.  Wightman functions turn out to be analytic with respect to
the commutative variables in a domain that is a two-dimensional
analogue of the corresponding analyticity domain in the standard
theory [3], [26]. This conclusion is based on the  $SO\,(1,1)$
invariance of the theory with respect to the commutative
variables. Besides, instead of the standard spectrality condition
[1], [3], only its analogue for the momentum components
corresponding to the commutative variables is used.  The
derivation of noncommutative analogues of the main results in the
axiomatic approach such as the CPT theorem and the spin–statistics
theorem is based on the analogue of local commutativity [28]
\be\label{2} [\vfhx, \vfhy] = 0, \ee 
if ${(x_0 - y_0)}^2 - {(x_3
-y_3)}^2 < 0$.

For the noncommutative variables, it suffices to assume that the
theory is invariant with respect to the reflection $I_s\,(2)$ for
noncommutative coordinates. Therefore, only the existence of the
$SO\,(1,1)\otimes I_s\,(2)$ symmetry and also of analogues of the
local commutativity and spectrality conditions is important here.
Moreover, some of the results are based only on the $SO\,(1,1)$
symmetry and on the spectrality condition.

In this report, we prove some classical results of axiomatic
quantum field theory for a real scalar field. We prove the
irreducibility of the set of field operators, the equivalence
between the local commutativity and symmetry conditions for
Wightman functions with respect to the permutation of arguments in
the analyticity domain. We prove the CPT theorem and introduce the
Borchers equivalence classes. Some similar results related to the
construction of Borchers equivalence classes were obtained also in
\cite{Fr}. We consider the theorem on the spin–statistics relation
as well.

We emphasize here that noncommutative analogues for the local
commutativity and spectrality conditions are also valid in the
commutative theory. It thus turns out that to reproduce the main
results of the axiomatic approach in the commutative theory, it
suffices to proceed from some assumptions that are weaker than the
standard ones. Noncommutativity plays the role of a new physical
reality making these weakenings of the local commutativity and
spectrality conditions inevitable. Of course, the results obtained
in this paper also remain true in other generalizations of the
standard theory that preserve the validity of the above-mentioned
analogues of the local commutativity and spectrality conditions.

This report has the following structure. Because the suggested
approach is a natural generalization of the standard formalism, we
begin with a presentation of the essence of this formalism. We
then introduce the Wightman functions in noncommutative QFT.

We first show that the translation invariance in respect to $\xz$
coordinate only together with the assumption that the set of basic
vectors contains only the positive energy vectors leads to the
irreducibility of the set of field operators. This result is valid
also if $\thzi \neq 0$. We then demonstrate that if the set of
vectors in momentum space does not contains taxyons in respect to
commutative variables( noncommutative spectral condition) then
Wightman functions are analytical functions in respect to
commutative variables in the domain similar to tubes. $SO\,(1,1)$
invariance gives us a possibility to enlarge this domain of
analyticity (enlarged tubes). This domain contains the subset of
Jost points, in which noncommutative LCC (local commutativity
condition) is fulfilled. This fact leads to the possibility to use
standard axiomatic methods and obtain above-mentioned results.

\section{Main postulates}

We first recall the essence of the Wightman approach in quantum
field theory. To simplify the formulas, we state this approach as
if the field $\vfx$  were defined at a point. In reality, the
operators are given by the expressions
\be\label{vf} \vff \equiv \int\,\vfx\,\fsx\,d\,x, \ee where $\fsx$
is a test function [1]–[3]. It is assumed that the vacuum vector
$\Psz$ is cyclic or, in other words, that the vectors of the type
\be\label{vec} \vfxo\,\vfxd\,...\,\vfxn\,\Psz \ee form a basis in
the space $J$ under consideration.

We assume that the space $J$ is endowed with an inner product
$<\cdot, \cdot>$, which can be indefinite. The cyclicity condition
for the vacuum means that each vector in the space $J$ can be
approximated arbitrarily accurately by vectors of type (4). We do
not specify this condition because the only essential thing in
what follows is that for all vectors, the inner product can be
approximated arbitrarily accurately by a linear combination of
Wightman functions
\be\label{wf} \wf = \Pszl, \vfxip\Pszr. \ee It is clear that the
inner product of two basis vectors is a Wightman function. Passing
to the noncommutative space, we can introduce the functions
\be\label{wfts} \wf  = \Pszl, \vfxipts \,\Pszr, \ee where Wightman
functions are written in the commutative space with the
corresponding changes in the product of operators.

 As is known (see [12], [13]), the product of two arbitrary
operators  $\vfx$ and $\psx$ in the noncommutative theory is
replaced with the  $\star$-product (Moyal-type product)
\be\label{skp} \phi\,(\hat x)\,\psi\,(\hat x) \rightarrow
\phi\,(x)\,\star\,\psi(x), \ee
where
\be\label{star} \phi\,(x)\,\star\,\psi\,(x) =
\phi\,(x)\,e^{\frac{i}{2}\,\theta_{\mu\nu}\,
\frac{\overleftarrow{\partial}}{\partial
x^\mu}\,\frac{\overrightarrow{\partial}}{\partial
y^\nu}}\,\psi\,(y)|_{x=y}\ . \ee

It is clear that there is an essential ambiguity in the choice of
the corresponding generalization of the Moyal-type product for
different points. It is remarkable that the possibility of
extending the axiomatic approach to the noncommutative theory is
independent of the concrete form of this generalization provided
that it satisfies some general requirements. As in the usual case,
we assume that the vectors of the type
\be\label{vets} \vfxipts \,\Psz \ee form a set of basis vectors in
the space $J_{\tilde \star}$  under consideration and that $\Psz$
is a cyclic vector. The simplest way to choose $\ts$ is to set
$\ts = 1$ for different neighboring points. Indeed, this choice
was made in [26]. But we note that the usual operator product is
used in [26] at coincident points as well, which, strictly
speaking, corresponds to the commutative theory with the
$SO\,(1,1)\otimes SO\,(2)$ invariance. An alternative natural
choice of the  $\ts$-product is the direct generalization of the
Moyal-type product to different points [13], i.e.,  $\ts = \star$.
This choice corresponds to the Wightman functions introduced in
[27]. In this case, the noncommutativity is manifested not only at
coincident points but also in their neighborhood. We thus
automatically obtain the correct expression at coincident
neighboring points.

The important point is the choice of test functions. We assume
that as $\thzi = 0$, test functions in respect to commutative
variables are standard ones. Precisely we assume that $\fsx$ can
be represented in the form $\fsx = f_1\,(x_0, x_3)f_2\,(x_1,
x_2)$, where $f_1\,(x_0, x_3)$ corresponds to tempered
distributions and $f_2\,(x_1, x_2)$ gives rise one of Shilov-
Gelfand spaces  \cite{GSh}. The exact determination of this space
is unessential as only the above-mentioned decomposition of $\fsx$
is important. For simplicity we formally can consider a field
operator as an operator defined in a point as all our reasonings
are valid also for $\vff$ (see eq. (3)). It follows from
expression (9) that in noncommutative theory the role of field
operators plays the combination $\vfx\,\ts$. As we consider a real
field, then
$$
{[\vfx\,\ts]}^+ = \vfx\,\ts \ .
$$
It is convenient to denote $\vfx\,\ts$ as $\tvfx$. In new
notations
\be\label{wft} \wf = \Pszl, \tvfxp\,\Pszr. \ee

We once again stress that axiomatic  results are based on the
commutativity and spectrality conditions, not on a definite choice
of the $\ts$-product.

\section{Irreducibility of the Field Operators Set}

As in \cc irreducibility of the field operators set in
noncommutative theory is a consequence both of translation
invariance of the theory  and corresponding spectral conditions.
In fact it is sufficient (the same is true also in commutative
case) that a vacuum vector would be invariant under the
translations along $\xz$ axis and all basic vectors in momentum
space $P_n$ satisfy the condition
$$
P_0 \geq 0\ .
$$
This condition and completeness of the set of basic vectors in
momentum space leads to equality
\be\label{8} \int\,d\,a\,e^{i\,p\,a}\,\langle\Phi,
U\,(a)\,\Psi\rangle \neq 0, \quad \mbox{only if} \quad p_0 \geq 0\ ,
\ee $\Phi$ and $\Psi$ are arbitrary vectors, $U\,(a)$ is a
translation operator, $a$ is a translation along $\xz$.

Let us remind the criterion of irreducibility. As $\tvfx$ is an
unbounded
operator, a proper definition of irreducibility would be: \\
{\bf Definition. } The set of operators $\tvfxi$ is irreducible if
any bounded operator $A$, which commutes with all field operators
\be\label{aco} \left[ A, \tvfx \right] = 0, \qquad \forall \; x
\ee is the following:
\be\label{ade} A = C\,\identity, \qquad C \in \complessi, \ee
$\identity$ is an identical operator.

To prove the irreducibility of the set of operators in question we
consider the expression
$$
\langle \Psz, A\,U\,(a)\,\vf(\txo)\,\vf(\txd)...\vf(\txn)\,\Psz
\rangle.
$$
Using translation invariance of $\Psz$ and \con (\ref{aco}), we
can write the chain of equalities:
$$
\langle \Psz, A\,U\,(a)\,\vf(\txo)\,\vf(\txd)...\vf(\txn)\,\Psz
\rangle = \langle \Psz, \vf\,(\txo + a) \cdots \vf\,(\txn +
a)\,A\,\Psz \rangle =
$$
\be\label{11} \langle \vf\,(\txn) \cdots \vf\,(\txo), U\,(-
a)\,A\,\Psz \rangle, \ee where  $a$ is a translation along $\xz$
axis. Let us remind that $\tvfxi$ is a Hermitian operator. So
\be\label{12} \langle A^+\,\Psz,
U\,(a)\,\vf(\txo)\,...\vf(\txn)\,\Psz \rangle = \langle
\vf\,(\txn) \cdots \vf\,(\txo)\,\Psz, U\,(- a)\,A\,\Psz \rangle.
\ee Let us consider
\bn\label{13} \int\,d\,a\,e^{i\,p\,a}\,\langle
A^+\,\Psz, U\,(a)\,\vf(\txo)\,...\vf(\txn)\,\Psz \rangle =
\nonumber\\
\int\,d\,a\,e^{i\,p\,a}\, \langle \vf\,(\txn) \cdots \vf\,(\txo),
U\,(- a)\,A\,\Psz \rangle. \en In accordance with eq. (\ref{8})
the left part of eq. (\ref{13}) is not zero only if $p \in \bar
V_2^+$ and the right part is not zero only if $- p \in \bar
V_2^+$. So if we take vector $\vf\,(\txn) \cdots
\vf\,(\txo)\,\Psz$, which is the linear combination of all basic
vectors excluding vacuum one, then
\be\label{14} \int\,d\,a\,e^{i\,p\,a}\, \langle \vf\,(\txn) \cdots
\vf\,(\txo), U\,(- a)\,A\,\Psz \rangle = 0 \quad \forall \; p. \ee
Eq. (\ref{14}) implies that
\be\label{15} \langle \Phi, A\,\Psz \rangle = 0 \ee for any vector
$\Phi$, which is the linear combination of all basic vectors
excluding vacuum one.

So
\be\label{16} A\,\Psz = C\,\Psz, \qquad C \in \complessi. \ee From
eq. (\ref{16}) it follows immediately that
\be\label{17} A\,\vf(\txo)\,...\vf(\txn)\,\Psz  =
C\,\vf(\txo)\,...\vf(\txn)\,\Psz. \ee As $A$ is a bounded operator
eq. (\ref{17}) implies that
\be\label{18} A\,\Phi = C\,\Phi, \qquad \forall \; \Phi \in
J_{\ts}. \ee

As translation invariance is valid in \nc theory in the general
case $\thzi \neq 0$, we can obtain the same result for general
case as well.

\section{Analytical Properties of  Wightman Functions}

As is known in commutative case analyticity of Wightman functions
in the primitive domain of analyticity (tubes) is the consequence
of spectral condition and translation invariance only. The
extension of this domain (the Bargman--Hall--Wightman theorem) is
realized in accordance with $SO\,(1,3)$ invariance of the theory.
In noncommutative case we have the  similar picture, but all
assumptions concern commutative coordinates only. Thus in
noncommutative case Wightman functions are analytical functions in
respect to commutative coordinates. Let us stress that
noncommutative coordinates remain real.

Taking into account  translation invariance in respect to the
commutative coordinates, it is convenient to write Wightman
functions in the form
\begin{equation}\label{ewf}
\wf = \wfksx,
\end{equation}
where $X$ denotes the set of noncommutative variables and $x_i^1,
x_i^2, \; i = 1,\ldots n$ and $\xi_i = \{\xi_i^0, \xi_i^3\}$,
where $\xi_i^0 = x_i^0 - x_{i +1}^0, \xi_i^3 = x_i^3 - x_{i
+1}^3$ .

Let us formulate the spectral condition. We assume that basic
vectors in momentum space are time-like vectors in respect to
commutative coordinates, that is
\begin{equation}\label{imp}
\cpnz \geq |\cpnt|\ .
\end{equation}
This condition is convenient to rewrite  as $\cpn \in
{\bar{V}}^{+}_{2}$, where ${\bar{V}}^{+}_{2}$ is a set of vectors,
satisfying the condition $x^0 \geq |x^3|$. Let us recall that
usual spectral condition is $\cpn \in {\bar{V}}^{+}$, that is
$\cpnz \geq |\vcpn|$. From condition (\ref{imp}) it follows that
eq. (11) leads to the following property of Wightman functions
\begin{equation}\label{wfsp}
W\,(P_1, ..., P_{n -1}, X) = \frac{1}{(2\pi)^{{n-1}}}\,
\int\,e^{i\,P_j\,\xi_j}\,W\,(\xi_1, ..., \xi_{n -1}, X) \,d\,\xi_1
...d\,\xi_{n -1} = 0,
\end{equation}
if $P_j \not \in \bar V_2^+$. The proof of (\ref{wfsp}) is similar
with the proof in commutative case \cite{SW}, \cite{BLT}. As
\begin{equation}\label{wfspo}
W\,(\xi_1, ..., \xi_{n -1}, X) =  \frac{1}{(2\pi)^{{n-1}}}\,
\int\,e^{- i\,P_j\,\xi_j}\,W\,(P_1, ..., P_{n -1}, X)\,d\,P_1
...d\,P_{n -1}\ ,
\end{equation}
we obtain immediately that $\wfnx$ is an analytical function in
the ``tube" $\; \ctnmi$\ :
\begin{equation}\label{tube}
\nui \in \ctnmi, \quad \mbox{if} \quad  \nui = \ksi - i\,\eti, \;
\eti \in  V_2^+, \; \eti = \{\etiz,  \etit\}\ .
\end{equation}
Let us stress that noncommutative coordinates are always real.

In accordance with the Bargmann-Hall-Wightman theorem $\wfnx$ is
the analytical function in the domain $\ctn$
\begin{equation}\label{ctn}
\ctn = \cup_{\Lac}\,\Lac\,\ctnmi,
\end{equation}
where $\Lac \in S_c\,O\,(1, 1)$ is a two-dimensional analog of
complex Lorentz group. As it was mentioned above $\ctn$ contains
the real points of analyticity, satisfying the condition (2) which
implies that
\begin{equation}\label{wftxi}
\wftxi = \wftxio,
\end{equation}
where $\txit$ are noncommutative Jost points , that is they
satisfy the conditions
\begin{equation}\label{app}
{\left(\txitz - \txjz\right)}^2 - {\left(\txitt - \txjt\right)}^2
< 0, \quad \forall \; i, j\ .
\end{equation}

\section{Relation between the Symmetry of Wightman Functions 
and the
Local \linebreak Commutativity Condition}

As a consequence of local commutativity condition (29), we have
\bn\label{wftx} \wftx = W\,\pitx, \en where  $\pitx$ is an
arbitrary permutation of the points $\txit$. Condition (30) means
that the function  $\wfz, \; \zi = \xit + i\,\yi$, is symmetric
with respect to its arguments in the domain into which  $T_n$
passes under the permutation of the points  $\txit$  with
subsequent application of transformations belonging to
$S_c\,O\,(1,1)$. Hence, the local commutativity condition permits
extending the analyticity domain for Wightman functions further.

We show that, conversely, the local commutativity condition
follows from the symmetry of Wightman functions in the analyticity
domain. For this, it suffices to prove that
\bn\label{wfztx} \wfztx = \wfztxo, \en where  $\txit$ and $\txipo$
are the Jost points and $z_j \in T^-_n$. Indeed, we can pass to
the limit $Im\,z_j = 0$ in~(\ref{wfztx}). To prove (\ref{wfztx}),
we note that the symmetry of Wightman functions in $T^-_n$ implies
the relation
\bn\label{26} \wfzp = \wfzpo. \en

The desired relation is proved if there is a tuple of points
$z'_j$ and a transformation $\La \; \in S_c\,O\,(1,1)$ under which
the points $z'_j$  pass into  $z_j, j\neq i, i +1$,  and the
points $z'_i, z'_{i + 1}$  pass into  $\txit$ è $\txipo$. The
corresponding transformation is similar to the one existing in the
commutative theory (see Chap.~5 in~[3]).

Therefore, the local commutativity condition, as in the
commutative case, is equivalent to the symmetry of Wightman
functions in the analyticity domain. This equivalence implies the
impossibility of a simple generalization of the local
commutativity condition, namely, the following assertion holds:

If
\be\label{tlcca} [\tvfx, \tvfy] = 0, \quad \mbox{for} \quad {(x_0
- y_0)}^2 - {(x_3 - y_3)}^2 < - l^2\ , \ee and if the usual axioms
are given, then the local commutativity condition is satisfied.

Indeed, the proof of the analyticity of Wightman functions does
not use the local commutativity condition, and we have the same
set of Jost points in this case as before. After the consideration
of the Jost points satisfying condition (33), we use the
above-mentioned local commutativity condition to prove the
symmetry of Wightman functions under the permutation of arguments.
It remains to use the fact that the local commutativity is a
consequence of the symmetry of Wightman functions. We note that
the similar result was also proved in the commutative case under
some more general assumptions \cite{Vlad}.

\section{ CPT Theorem and Borchers Equivalence Classes}

By definition, the CPT-transform operator for real field is given
by the formula
\be\label{Th} \Theta\,\tvf\,\Theta^{- 1} = \tvfmx. \ee According
to the Wigner theorem (see  \cite{SW}), the constructed theory is
CPT-invariant if $\Theta$  is antiunitary, i.e., if
$$
\langle \Th\,\Phi, \Th\,\Psi \rangle  = \langle \Psi, \Phi \rangle
\quad \forall \; \Psi \; \mbox{and} \; \Phi.
$$
Because all inner products of basis vectors (9) reduce to Wightman
functions, it suffices to find a condition under which ì
\be\label{Thñ} \langle \Th\,\Psz, \Th\,\tvfxp\,\Psz \rangle =
 \overline {\langle \Psz, \tvfxp\,\Psz \rangle},
\ee where the overbar denotes complex conjugation. We can easily
show that (35) holds if
\be\label{30} \langle \Psz, \tvfxp\,\Psz \rangle = \langle \Psz,
\tvfmxpo\,\Psz \rangle. \ee According to the CPT theorem, relation
(\ref{30}) is equivalent to the weak local commutativity condition
\be\label{31} \langle \Psz, \tvftxo\, ...\tvftxn\,\Psz \rangle =
\langle \Psz, \tvftxn\, ...\tvftxo\,\Psz \rangle,  \quad \txit
\sim \txj\ . \ee Condition (37) is obviously satisfied if the local
commutativity condition holds. But we stress that there are some
well-known examples of fields  \cite{BLT} that satisfy the weak
local commutativity condition but do not satisfy the usual local
commutativity condition. Because the equivalence of  (\ref{30})
and (\ref{31}) has been proved \cite{LAG, CMNTV, MV}, we do not
dwell on it. We only note that in this case, the $I_s\,(2)$
invariance of the theory with respect to the noncommutative
coordinates is used essentially. In the usual situation, the
invariance of Wightman functions under complex Lorentz
transformations includes the invariance under the change of sign
for all coordinates. In the noncommutative case, the
$S_c\,O\,(1,1)$  invariance means the related invariance only
under the change of sign for the commutative coordinates. The
invariance of Wightman functions under the change of sign of
noncommutative coordinates follows from the  $I_s\,(2)$
invariance.

Let us point out that CPT theorem is valid in any $S\,O\,(1,1)
\otimes  I_s\,(2)$ invariant theory if only Wightman functions are
analytical ones in the above-mentioned domain.

We now establish conditions under which two operators  $\tvfx$ and
$\tpsx \equiv \psx\,\ts$ have a common CPTtransform operator. We
show that in this situation as well as in the commutative case,
the above-mentioned property holds under the condition that
\bn\label{32} \langle \Psz, \tvftxo\,
...\tvftxi\,\tpsx\,\tvftxipo\, ...,\tvftxn\,
\Psz \rangle =     \nonumber \\
\qquad {} \langle \Psz, \tvftxn\, ...\tvftxipo\,\tpsx\,\tvftxi\,
...\tvftxo\,\Psz \rangle, \en if $\txit \sim \txj,\; x \sim \txj
\; \forall \; i,j$. We assume here that the field $\tpsx$
satisfies the conditions under which the domain of analyticity of
the functions in (\ref{32}) is the same as that for the Wightman
function of the field $\tvfx$, but $\Psz$ need not be a cyclic
vector for the fields $\tpsx$. Moreover, we assume that the
$I_s\,(2)$ invariance with respect to the noncommutative variables
holds.

In this case, by analogy with the implication under which
(\ref{30}) follows from (\ref{31}), the relation
\bn\label{33} \langle \Psz, \tvfxo\, ...\tvfxi\,\tpsx\,\tvfxipo\,
...,\tvfxn\,
\Psz \rangle =    \nonumber \\
\langle \Psz, \tvfmxn\, ...\tvfmxipo\,\tpsmx\,\tvfmxi\,
...\tvfmxo\,\Psz \rangle \en is implied by condition (\ref{32})
for arbitrary  $x, \xo, ... \xn$. It can be easily seen that
(\ref{33}) can be written in the form
\be\label{34} \langle \Phi, \tpsx\,\Psi \rangle = \langle
\Th\,\Psi, \tpsmx\,\Th\,\Phi \rangle, \ee where $\Phi =
\tvfxi...\tvfxo\,\Psz, \: \Psi = \tvfxipo...\tvfxn\,\Psz$.

It is easy see that by the antiunitarity of the operator $\Th$, we
always have
\bn\label{35} \langle \Phi, \tpsx\,\Psi \rangle = \langle
\Th\,\Psi, \Th\,\tpsx\,{\Th}^{- 1}\,\Th\,\Phi  \rangle. \en
Comparing (\ref{34}) and (\ref{35}) gives the desired result
because  $\Phi$ and  $\Psi$ are arbitrary basis vectors. In turn,
condition (\ref{33}), as in the case of the CPT theorem, leads to
relation (\ref{32}) . Furthermore, because  $\Th$ is antiunitary,
$\tpsx$ is a weakly local field, i.e., a field satisfying the weak
local commutativity condition. Moreover, the condition of mutual
weak local commutativity holds here. This means that the Wightman
function containing arbitrarily many fields  $\tvfxi$ and $\tpsyj$
does not change at the Jost points under the replacement of the
direct order of the arguments by the reverse order.

The most important conclusion is that the mutual weak locality
condition is transitive, i.e., if each of the fields  $\tpsox$ and
$\tpsdx$ is weakly local in relation to a field  $\tvfx$, then
they are mutually weakly local as well. Indeed, according to the
previous argument, their CPT-transform operators coincide.

Hence, weakly local fields, as in the commutative case, form a
Borchers equivalence class that includes the field $\tvfx$ and
also all fields each of which is weakly local in relation to
$\tvfx$, and these fields turn out to be weakly local between
themselves in this case. The most important property of these
fields is that the $S$-matrices for the fields belonging to a
common Borchers equivalence class coincide if the limits of the
fields as $\xz \to - \infty$ or $\xz \to + \infty$ coincide.
Indeed, as is known, the $S$-matrix is related to the
CPT-transform operators by  \cite{BLT} \pagebreak

\be\label{36} S =  {\Th}^{- 1}\,{\Th}_{out} = {\Th}^{-
1}_{in}\,\Th, \ee where  ${\Th}_{in}$ and ${\Th}_{out}$  are the
corresponding operators for the asymptotic fields.

It can be shown that the local commutativity condition is also
transitive. This means that two fields each of which is local in
relation to a field $\tvfx$ for which $\Psz$ is a cyclic vector
are mutually local. The proof of this property is similar to the
proof in the commutative case \cite{BLT}.

\section{ Spin-statistics Theorem }

The theorem on the spin–statistics relation for a complex scalar
field can be reduced to the corresponding theorem for a Hermitian
field using the following lemma.\\[-4mm]

{\large \bf Lemma } 

 If the commutation relations
\be\label{tco} [\tvfx, \tpsy] = 0,  \qquad x \sim y, \ee
\be\label{tfco} \{ \tvfx, \tpscy \} = 0,  \qquad x \sim y. \ee
hold and if $\tvfx \neq 0$, then $\tpsy \equiv 0$.

As in the usual case, we assume that the operators $\tvfx$ and
$\tpsy$ act in a subspace with a positive definite metric.

The proof of the Lemma essentially uses the cluster properties of
Wightman functions. In view of local commutativity condition
(\ref{wftxi}), the relations
\bn\label{39} \langle \Psz, \tvfxo\, ...\tvfxi\,\tvfxipopla\,
...,\tvfxnpla\,
\Psz \rangle \rightarrow   \nonumber \\
\langle \Psz, \tvfxo\, ...\tvfxi\,\Psz \rangle \langle \Psz,
\tvfxipo\, ...\tvfxn\,\Psz \rangle, \en are sure to hold if $\la
\to \infty, \: a_0^2 - a_3^2 = - 1$. The proof of (\ref{39})
reduces to the corresponding proof in the commutative case
\cite{SW}.

To prove the lemma, it suffices to note that according to
(\ref{tco}) and  (\ref{tfco}), we have
$$
\langle \Psz, \tvfcx\,\tpscy\,\tpsy\,\tvfx\,\Psz \rangle = -
\,\langle \Psz,\tvfcx\,\tvfx\,\tpscy\,\tpsy\,\Psz \rangle
\rightarrow
$$
\bn\label{40} - \,\langle \Psz, \tvfcx\,\tvfx\,\Psz \rangle
\langle \Psz, \tpscy\,\tpsy\,\Psz \rangle, \en for  $x \sim y$ if
$y = x + \la\,a,  \: \la \to \infty, \: a_0^2 - a_3^2 = - 1$.
Noting that the original Wightman function is positive and that
the third expression in (\ref{40}) is negative, we conclude that
one of the relations
\be\label{41} \tvfx\,\Psz = 0 \quad \mbox{or } \quad \tpsy\,\Psz =
0. \ee holds. We show that the first relation in (\ref{41}) means
that $\tvfx \equiv 0$ and the other implies that  $\tpsy \equiv
0$. For simplicity we consider the proof for real field $\tvfx$,
our arguments can be easily extended to a complex field. Indeed,
for instance, let the second relation in (\ref{41}) hold. In view
of (\ref{tco})  and (\ref{tfco}), we then have
\be\label{42} \langle \Psz, \tvftxo\,
...\tvftxi\,\tpsy\,\tvftxipo\, ...,\tvftxn\, \Psz \rangle =  0,
\ee if  $y \sim \txj, \; \forall \; j$. Because vanishing of the
Wightman function at the Jost points means that it is identically
zero, we have
\be\label{43} \langle \Psz, \tvfxo\, ...\tvfxi\,\tpsy\,\tvfxipo\,
...,\tvfxn\, \Psz \rangle =  0,   \quad \forall \; \xit  \ee for
all $\xit$. Because  $\Psz$ is a cyclic vector for the field
$\tvfx$, we conclude that $\tpsy \equiv 0$. The same argument also
applies to  $\tvfx$. By assumption, $\tvfx \neq 0$ and the
assertion of the lemma is thus proved.

We next apply the lemma to the case $\tpsx = \tvfx$ and prove that
if
\be\label{44} \{ \tvfx, \tvfcy \} = 0,  \qquad x \sim y, \ee then
$\tvfx \equiv 0$ then $\tvfx \equiv 0$. For this, we consider the
functions
$$
W_1\,(x - y) = \langle \Psz, \tvfx\,\tvfcy\,\Psz \rangle,
$$
$$
W_2\,(x - y) = \langle \Psz, \tvfcx\,\tvfy\,\Psz \rangle.
$$
Using the argument presented in \cite{CMNTV} for these functions,
we conclude that (\ref{44}) implies that  $\tvfx\,\Psz = 0$, i.e.
$\tvfx \equiv 0$.
\vspace*{-1mm}
\section{Conclusion}
\vspace*{-2mm}
The presented investigation shows that whereas there are analogues
of local commutativity~(\ref{wftxi}) and spectrality
(\ref{wfsp}), the $SO\,(1,1)\otimes I_s\,(2)$ invariance of the
theory suffices for deriving the main results in the axiomatic
approach. A physically interesting version of the latter theory is
the noncommutative quantum field theory in which time commutes
with the spatial coordinates.\\

\end{document}